\documentclass[10pt,letterpaper]{article}
\usepackage{arydshln}
\usepackage{authblk}
\usepackage[utf8]{inputenc}
\usepackage[backend=biber,style=numeric-comp,sortcites=true,minnames=6,maxbibnames=7,maxcitenames=7,giveninits=true,terseinits=true,sorting=none]{biblatex}
\usepackage{caption}
\usepackage[a4paper]{geometry}
\usepackage{graphicx}
\graphicspath{{Figures/}}
\usepackage{multirow}
\usepackage[pagewise]{lineno}
\usepackage{subcaption}
\usepackage{todonotes}
\usepackage{ulem}
\usepackage{verbatim}
\usepackage{wasysym}
\usepackage{todonotes}

\bibliography{bibliography.bib}

\DeclareFieldFormat[book, article, thesis, inproceedings]{title}{#1}
\DeclareFieldFormat{journaltitle}{#1}
\DeclareFieldFormat{booktitle}{#1}
\DeclareFieldFormat{postnote}{\mkcomprange[{\mkpageprefix[pagination]}]{#1}}
\DeclareFieldFormat{pages}{\mkcomprange{#1}}

\DeclareNameAlias{default}{family-given}

\renewbibmacro{in:}{\ifentrytype{article} {} {\printtext{\bibstring{in} \intitlepunct}}}

\renewbibmacro*{journal+issuetitle}{%
  \usebibmacro{journal}%
  \setunit*{\addperiod\addspace}%
  \iffieldundef{series}
    {}
    {\newunit
     \printfield{series}%
     \setunit{\addspace}}%
  \setunit{\addperiod\addspace}%
  \usebibmacro{issue+date}%
  \setunit{\addcolon\space}%
  \usebibmacro{issue}%
  \setunit{\addsemicolon\space}%
  \usebibmacro{volume+number+eid}%
  \newunit}
  
\DeclareFieldFormat[article,periodical]{number}{\mkbibparens{#1}}
\renewbibmacro*{volume+number+eid}{%
  \printfield{volume}%
  \printfield{number}%
  \setunit{\addcomma\space}%
  \printfield{eid}}

\renewbibmacro*{issue+date}{%
  \iffieldundef{issue}
    {\usebibmacro{date}}
    {\printfield{issue}%
     \setunit*{\addspace}%
     \usebibmacro{date}}%
  \newunit}

\makeatletter
\renewcommand{\maketitle}{\bgroup\setlength{\parindent}{0pt}
\begin{flushleft}
  \Large{\textbf{\@title}}
  
  \vspace{2ex}
  \textit{\normalsize{\@author}}
\end{flushleft}
\egroup
}

\makeatother

\title{Optically stimulated luminescence system as an alternative for radiochromic film for 2D reference dosimetry in UHDR electron beams.}
\author[a,b,c,$\dagger$]{\underline{Vanreusel Verdi}}
\author[b,c]{Gasparini Alessia}
\author[d]{Galante Federica}
\author[d]{Mariani Giulia}
\author[d]{Pacitti Matteo}
\author[b]{Colijn Arnaud}
\author[e]{Reniers Brigitte}
\author[e]{Yalvac Burak}
\author[f]{Vandenbroucke Dirk}
\author[g]{Peeters Marc}
\author[f]{Leblans Paul}
\author[d]{Felici Giuseppe}
\author[a]{de Freitas Nascimento Luana}
\author[b,c]{Verellen Dirk}
\affil[a]{Research in Dosimetric Applications, SCK CEN, Boeretang 200, 2400 Mol, Belgium}
\affil[b]{CORE, University of Antwerp, Universiteitsplein 1, 2610 Wilrijk, Belgium}
\affil[c]{Iridium Netwerk, Oosterveldlaan 22, 2610 Wilrijk, Belgium}
\affil[d]{Sordina IORT Technologies S.p.A., Via dell’Industria, 1/A, 04011 Aprilia, Latina, Italy}
\affil[e]{NuTeC, CMK, Hasselt University, Wetenschapspark 27, 3590 Diepenbeek, Belgium}
\affil[f]{Agfa N.V., Septestraat 27, 2640 Mortsel, Belgium}
\affil[g]{UZA, Edegem, Belgium}

\begin{document}
\maketitle
\vspace{5ex}
\section*{Abstract} 
Radiotherapy is part of the treatment of over 50\% of cancer patients. Its efficacy is limited by the radiotoxicity to the healthy tissue. FLASH-RT is based on the biological effect that ultra-high
dose rates (UHDR) and very short treatment times strongly reduce normal tissue toxicity, while
preserving the anti-tumoral effect.
Despite many positive preclinical results, the translation of FLASH-RT to the clinic is hampered by the lack of accurate dosimetry for UHDR beams. To date radiochromic film is commonly used for dose assessment but has the drawback of lengthy and cumbersome read out procedures.
In this work, we investigate the equivalence of a 2D OSL system to radiochromic film dosimetry in terms of dose rate independency. 
The comparison of both systems was done using the ElectronFlash linac. We investigated the dose rate dependence by variation of the 1) modality, 2) pulse repetition frequency, 3) pulse length and 4) source to surface distance. Additionally, we compared the 2D characteristics by field size measurements.
The OSL calibration is transferable between conventional and UHDR modality. Both systems are equally independent of average dose rate, pulse length and instantaneous dose rate. The OSL system showed equivalent in field size determination within 3 sigma.
We show the promising nature of the 2D OSL system to serve as alternative for radiochromic film in UHDR electron beams. However, more in depth characterization is needed to assess its full potential. 

\vfill

\flushleft{$^\dagger$\textit{Corresponding author:} verdi.vanreusel@sckcen.be \newline \textit{Address:} Boeretang 200, 2400 Mol, Belgium}\\\vspace{2ex}
\textit{\textbf{Keywords:}} FLASH-RT, Reference Dosimetry, Optically Stimulated Luminescence
\thispagestyle{empty}

\newpage
\setcounter{page}{1}
\section{Introduction}
In 2014 Favaudon et al. brought renewed interest to the use of ultra-high dose rates (UHDR) in radiation therapy. In their paper, they describe the so-called FLASH-effect. This is a biological effect, where the combination of UHDR with very short treatment times shows an increased differential effect between normal tissue and tumor cells. Namely, they showed a strong sparing of normal tissue with conservation of the anti-tumor effect for lung tumor therapy in mice~\cite{favaudon2014ultrahigh}. Since then, many other preclinical studies have reproduced the sparing effect of these UHDR irradiations in various organs, using different beam types~\cite{montay2017irradiation, montay2018x, montay2019long, simmons2019reduced, vozenin2019advantage, vozenin2019biological, alaghband2020neuroprotection, fouillade2020flash, soto2020flash,  sorensen2022vivo, tinganelli2022flash}. Also, a dozen of studies have validated the preservation of the anti-tumor effect~\cite{cunningham2021flash, montay2021hypofractionated, velalopoulou2021flash, eggold2022abdominopelvic, rohrer2022dose}. 
Despite its promising nature, FLASH-RT also comes with challenges that hamper its clinical translation. One of the major challenges is dosimetry, as the state-of-the-art dosimeters for conventional radiation therapy are subject to saturation effects at these UHDR~\cite{jorge2019dosimetric, di2020flash, mcmanus2020challenge, wilson2020ultra,szpala2021dosimetry, wu2021technological}. 
\newline
Many FLASH studies rely on radiochromic film for dose assessment as it is currently the only solution for 2D UHDR dosimetry. The dose rate independence is guaranteed up to several Gy per minute, while expected to extend to higher dose rates. The dose uncertainty of radiochromic film is, however, in the order of 5\%, and readout procedures are cumbersome and lengthy. It has a non-linear dose-response relationship, and the AAPM guidelines recommend to scan the radiochromic film only 16-24 hours post irradiation~\cite{niroomand2020report}. Therefore, we investigated a 2D optically stimulated luminescence (OSL) system as alternative for 2D UHDR dosimetry.
\newline
In luminescence dosimetry, OSL is an established technique with applications in several fields such as conventional radiotherapy, personal dosimetry, space dosimetry, etc.~\cite{yukihara2008optically,sommer2011new,yukihara2022luminescence}. As an example, by coupling the OSL material with an optical fiber system, OSL dosimetry can be used for in-vivo dose assessment. However, the current applications, including this example, are point measurements. While meeting the needs for personal, space and accidental dosimetry, radiotherapy applications would benefit from 2D dose assessment. It is indispensable for, for example intensity modulation and image guidance therapies.
\newline
More recent studies have shown the potential of 2D luminescence dosimetry~\cite{olko2006new,de2022investigation}. Recent publications have showed the potential of BaFBr OSL phosphor sheets in conventional MV photon beams with reproducible dose assessment with an accuracy of 3\%~\cite{wouter2017reusable,de2019extended}. One of the challenges associated with BaFBr OSL material is the dark decay. This is the effect where the OSL signal fades over time due to spontaneous recombination of the electron-hole pairs. A mitigation for this has recently been proposed by Caprioli et al, who developed a model to correct the signal for the dark decay independently from the irradiation process~\cite{caprioli2022calibration}. The combination of these studies makes the OSL system an easy and accurate 2D dosimetry tool in conventional radiotherapy beams. 
\newline
As long as not all charge traps in the phosphor are filled, no dose rate dependence is expected. Therefore, this study characterises a 2D BaFBr-based OSL system, and investigates whether it could serve as a more convenient and faster alternative to radiochromic films in both conventional and UHDR MeV electron beams.
\newline

\section{Materials and Methods}
The dose rate dependency of an OSL system in UHDR electron beams was investigated by variation of the 1) pulse repetition frequency (PRF); 2) pulse duration; 3) dose rate modality; and 4) source-surface distance (SSD). The OSL system was benchmarked with radiochromic film and a flashDiamond detector at different dose rates. It was opted to use 2 benchmark dosimeters as their behavior with varying dose rate has only been sparsely investigated in the literature~\cite{marinelli2022design,verona2022application}. By using 2 dosimeters based on different working principles, the dose assessment can be assumed correct when they are both in agreement. In case of disagreement, a third benchmark dosimeter is included with alanine EPR dosimetry.   

\subsection{Dosimeters}
\subsubsection*{Optically stimulated luminescence system}
\begin{figure}[!ht]
    \centering
    \begin{subfigure}{.45\textwidth}
    \centering
        \includegraphics[width=\textwidth]{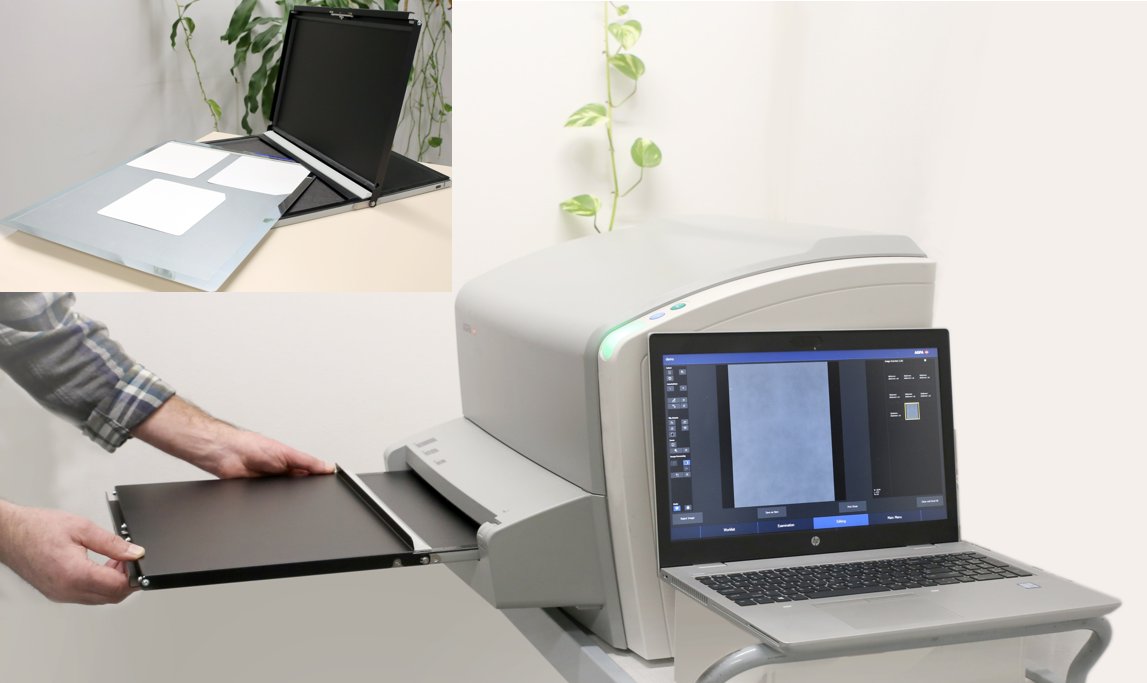}
        \caption{\label{fig: crreader}}
    \end{subfigure}
    \begin{subfigure}{.39\textwidth}
    \centering
        \includegraphics[width=\textwidth]{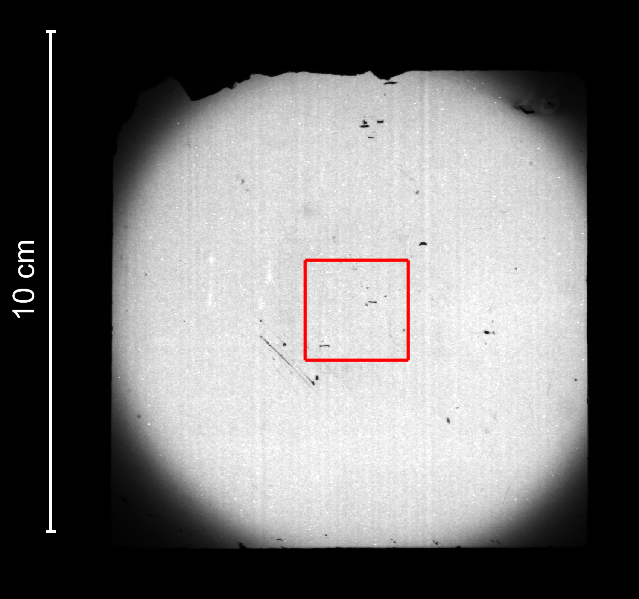}
        \caption{\label{fig: roiextraction}}
    \end{subfigure}
    \caption{CR reader with light tight cassette to transport the OSL sheet (a), and a response map with a red frame, showing the region of interest which is automatically determined (b).}
\end{figure}

Upon irradiation with ionizing radiation, electron-hole pairs are created and stored in storage phosphor crystal defects. The trapped electrons can be liberated by means of optical stimulation. The electrons will recombine with trapped holes leading to the emission of light of which the intensity is proportional to the absorbed dose. An OSL system is expected to be a good candidate for UHDR dosimetry as its working principle is not based on charge collection and the associated problems of ion recombination or overflow of electronics. As long as not all charge traps in the phosphor are filled, no dose rate dependence is expected. 
\newline
A semi-flexible sheet was coated with a dedicated phosphor of BaF(Br,I):Eu$^{2+}$ with small grain size (in the micrometer range) (table~\ref{tab: OSL Sheet description}). Monte Carlo simulations suggest that a small grain size leads to a reduces the energy dependence that was observed in~\cite{wouter2017reusable}. Readout was done with a CR-15 (Agfa N.V., Mortsel, Belgium) computed radiography (CR) scanner (Fig.~\ref{fig: crreader})~\cite{crijns2015t}. The scanner has a 80 mW solid state laser, emitting at ca. 665 for stimulation and a photo multiplier tube (PMT) to detect, amplify and digitize the light signal.  The laser light is scanned over the sheet in the x-direction by a scanning mirror and in the y-direction by physical movement of the sheet. This way the stimulated emission is obtained point by point which allows 2D digitization.  Since the OSL sheet is stimulated by light in the visible spectrum, irradiations were performed in dark and the OSL sheet was put on a carrier which fits in a light tight cassette for transport. Induced light fading was avoided by automatic insertion of the carrier from the light tight cassette into the CR scanner. The effect of dark fading was excluded by keeping the time between start of irradiation and readout constant at 117.6 s ± 3.0 s. Both coating of the sheets and the development of the CR-reader, cassette, carrier system and readout software were done by Agfa N.V. (Mortsel, Belgium)~\cite{leblans2011storage,Agfa2023CR} within the QUARTEL [HBC.2020.3003] and eFLASH2D [HBC.2021.0946] projects funded by VLAIO (Medvia). Several settings, such as the PMT voltage, were decreased to allow dose readings up to 150~Gy. Two images were acquired at the start of each measurement day to warm up the electronics and prevent signal drift. The OSL sheet was erased and read out to guarantee the absence of a ghost image.  

\subsubsection*{Radiochromic film}
EBT-XD films from Ashland (Wilmington, USA) were used for comparison (table~\ref{tab: OSL Sheet description}). The radiochromic film was scanned 3 times, 2 days post irradiation in transmission mode with an Epson 10000XL scanner (Epson, Japan). A scan resolution of 150 dpi, resulting in 0.163x0.163 mm$^2$ pixels, was chosen, except for the films that were used to determine the beam profile. For these films a resolution of 254 dpi was chosen such that the pixel size matches the one of the OSL sheet (0.1x0.1 mm$^2$). Pixel values were translated to dose via an in-house MATLAB (MathWorks, United States). The 3 scans were averaged, after which dose was obtained by single channel dosimetry. Every measurement day, a set of films was irradiated simultaneously with an Advanced Markus (AM) plane parallel ionization chamber (PTW, Freiburg, Germany) in conventional modality to serve as reference. The color channel of which the dose measurement showed the best agreement with the AM reading was chosen for further dose calculations. This was the green color channel for all measurement days.
\newline
The dose was averaged over a square region of interest in the center of the field with a width of 150 pixels.

\subsubsection*{flashDiamond}
A prototype of the flashDiamond dosimeter (PTW, Freiburg, Germany) was used as independent reference dosimeter (table~\ref{tab: OSL Sheet description}) \cite{marinelli2022design, verona2022application}. A flashAdapter was used to decrease the instantaneous output  current, and warrant compatibility with a Unidos Webline electrometer. 
\subsubsection*{Alanine pellets}
Alanine electron paramagnetic resonance (EPR) dosimetry (NuTeC, Hasselt, Belgium) was used when a discrepancy was observed between radiochromic film and flashDiamond dose readings. The alanine pellets were cylindrical with a diameter of 5 mm and a thickness of 2.8 mm. Read out was performed at NuTeC (UHasselt) using a Bruker EMXmicro spectrometer with a 9-inch magnet, equipped with a high-sensitivity resonator ER4119HS-W1. The alanine pellets are traceable to the primary standard at PTB (Braunschweig, Germany).

\begin{table}[!ht]
    \centering
    \caption{The used dosimeters.}
    \resizebox{\textwidth}{!}{ 
    \begin{tabular}{ccccl}
         Dosimeter & Dimensionality & Dimensions [mm$^3$] & pixel size [$\mu$m$^2$]& Dosimeter specific characteristics \\
         \hline
         \hline \\[-1ex]
         \multirow{4}{*}{BaF(Br,I):Eu$^{2+}$ OSL} & \multirow{4}{*}{2D} & \multirow{4}{*}{100x100x0.2} & \multirow{4}{*}{100x100} & Average crystal size = 1.5 $\mu$m\\
         &&&& Concentration = 5.3\%\\ 
         &&&& Coating thickness = 8 $\mu$m\\
         &&&& Coating weight = 0.5 g/m$^2$ \\
         &&&&\\
         \multirow{3}{*}{radiochromic film} & \multirow{3}{*}{2D} & \multirow{3}{*}{40x50x0.275} & \multirow{3}{*}{169.3x169.3} & film type = EBT XD\\
         &&&& single channel dosimetry\\ 
         &&&& Color channel for dose conversion = green\\
         &&&&\\
         \multirow{2}{*}{flashDiamond prototype} & point & $\diameter$ = 1.1 mm; & \multirow{2}{*}{N.A.} & \multirow{2}{*}{Serial number =7606} \\
         &dosimeter&thickness = 1 $\mu$m&&\\
         &&&&\\
         \multirow{2}{*}{Alanine EPR} & point & $\diameter$ = 5 mm; & \multirow{2}{*}{N.A.} &  \\
         &dosimeter&thickness = 2.8 mm&&\\
    \end{tabular}}
    \label{tab: OSL Sheet description}
\end{table}

\subsection{Setup}
\begin{figure}[!ht]
    \centering
    \includegraphics[width=.5\textwidth]{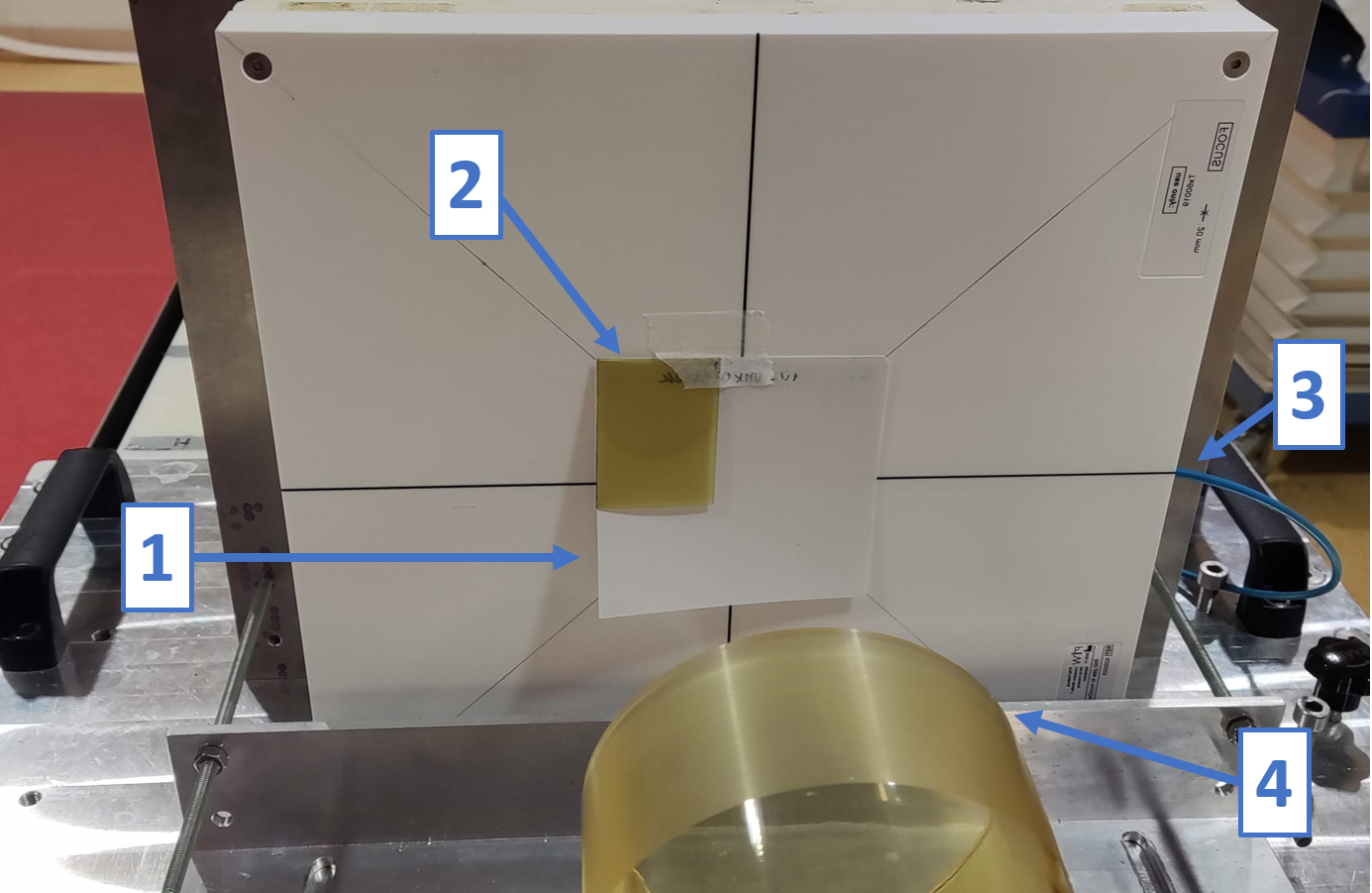}
    \caption{Setup without buildup slabs with the OSL sheet (1) and a radiochromic film (2) in front of a dedicated slab that contains a flashDiamond dosimeter (3), aligned with the center of the beam. The applicator of the linac is denoted by (4).}
    \label{fig: SetUp}
\end{figure}

The OSL sheet was positioned at depth of maximum dose deposition in a RW3 (PTW) phantom (15 mm). The flashDiamond was irradiated simultaneously and placed in a dedicated RW3 slab with the effective depth of measurement at 20 mm. The alanine pellet was set in a custom made RW3 slab, right in front of- and in contact with the OSL sheet. Figure~\ref{fig: SetUp} shows a typical setup, excluding buildup slabs, with an OSL sheet (1), radiochromic film (2), and a dedicated slab containing the flashDiamond dosimeter (arrow 3 points at the cable where it enters the phantom). Care was taken that the sheet was flat when sandwiched between the buildup and flashDiamond slabs. Except for experiments with various SSDs, the RW3 phantom was put in contact with the PMMA applicator of the linac (4). The phantom was put on a sledge to allow reproducible variation of the SSD.

\subsection{Irradiations}
\begin{figure}[!ht]
    \centering
    \begin{subfigure}{.45\textwidth}
    \centering
        \includegraphics[width=\textwidth]{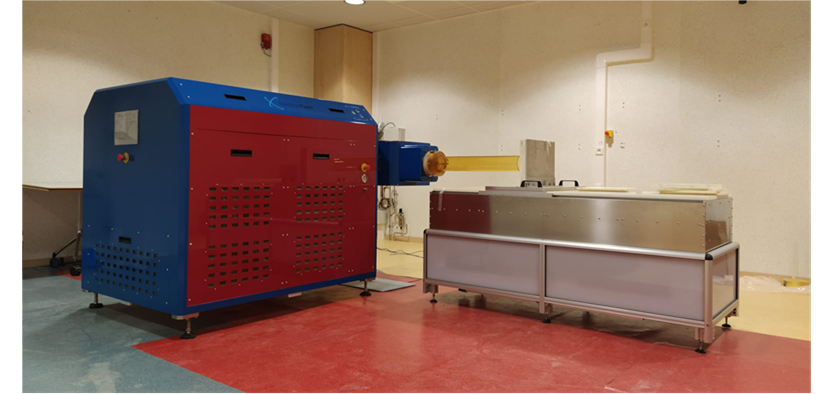}
        \caption{\label{fig: ElectronFlash}}
    \end{subfigure}
    \begin{subfigure}{.45\textwidth}
    \centering
        \includegraphics[width=\textwidth]{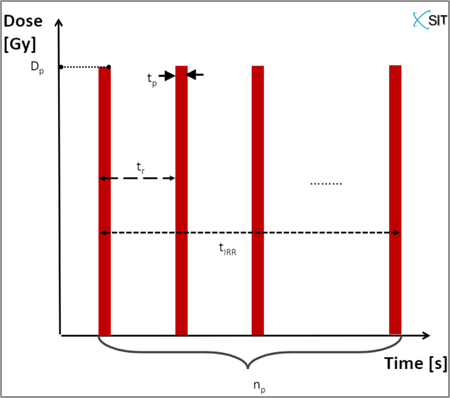}
        \caption{\label{fig: PulseScheme}}
    \end{subfigure}
    \caption{The ElectronFlash linac (a), and the typical pulse shape of an electron beam, with the parameters that are considered important to obtain the FLASH effect (b). D$_p$ denotes the pulse amplitude, t$_{p}$ the pulse length, t$_r$ the time between the pulses, t$_{IRR}$ the total treatment time, and n$_p$ the number of pulses.}
\end{figure}

All irradiations were performed using the ElectronFlash linac (SIT, Aprilia, Italy), shown in~figure~\ref{fig: ElectronFlash}. This linac was developed for preclinical FLASH research with electron beams, with nominal energies of 7 and 9 MeV~\cite{di2020flash}. Electron beams are delivered in a pulsed manner as represented in figure~\ref{fig: PulseScheme}. The ElectronFlash allows systematic variation of several beam parameters that are considered important to obtain the FLASH-effect~\cite{vozenin2020all}.  Pulse amplitude can be altered by switching between conventional and UHDR modalities; Pulse length can be varied in UHDR modality between 0.5 and 4 $\mu$s; and Pulse repetition frequency (PRF) can be varied between 1 and 245 Hz for the longest pulse length and can range up to 350 Hz for shorter pulse lengths. 
\newline
The dose rate was studied by varying the following parameters:
\begin{enumerate}
    \item the pulse amplitude/instantaneous dose rate (D$_p$) by comparing conventional and UHDR modalities
    \item the PRF (1/t$_r$)
    \item the pulse length (t$_p$)
    \item the pulse amplitude/instantaneous dose rate (D$_p$) by changing the SSD.
\end{enumerate}
The dose rate can be described in terms of these variables (visualized in figure~\ref{fig: PulseScheme}) as:
\begin{equation}
    \dot{\textrm{D}} = \frac{\textrm{D}_p\cdot\textrm{t}_p}{\textrm{t}_r}.
\end{equation}
The experiment comparing conventional and UHDR modalities, was also used to calibrate the OSL system against radiochromic film and the flashDiamond in both modalities. Circular fields with diameters of 10 and 3.5 cm were used for UHDR and conventional irradiations respectively. The beam parameters used for all experiments are tabulated in appendix~\ref{sec: appendix}. All irradiations were repeated 3 times with the OSL sheet and the flashDiamond in place. Radiochromic film was added for at least one of these irradiations. The alanine irradiation was performed as an additional measurement, in combination with the OSL sheet and the flashDiamond. 

\subsection{Processing}
The OSL response was stored as 16-bit dicom image with a spatial resolution of 100x100 $\mu$m$^2$. The average grey value of a square region of interest was extracted using an in-house MATLAB (MathWorks, United States) script. The center of the irradiated field was automatically determined and aligned with the center of the region of interest (ROI) as shown in figure~\ref{fig: roiextraction}. The size of the ROI was chosen such that the same ROI could be used for both the 10 and 3.5 cm diameter fields. If the automatic alignment failed, manual ROI positioning was performed. Background subtraction was performed, based on the background image acquired at the start of each measurement day. In the remainder of this paper, the response of the sheet will refer to the background subtracted average grey value of an image. 
\newline
Calibration of the OSL sheet was performed by linear fitting of the dose response curve against radiochromic film and flashdiamond doses, with the intercept forced to zero. For the subsequent experiments, the grey value was converted to dose by multiplication with the calibration factor obtained from the experiment comparing conventional and UHDR modalities.
\newline
To investigate the 2D characteristics of the OSL sheet in comparison with radiochromic film, the diameter of the field was measured for two field sizes. A diagonal profile was extracted from 2 images from the first experiment. These images were flat field corrected to allow a fair comparison. A flat field was obtained by irradiating the sheet at the largest SSD, at the end of the table, without applicator. The flat field image was manually registered with the field size dose map. The flat field image was normalized to the 99.5th percentile grey value. The field size dose map was then divided by the normalized flat field image. Hereafter, the field size dose maps obtained with the OSL system and radiochromic film were normalized. A band of 1.7 mm wide, centered around the center of the field was extracted from both dose maps to obtain the profiles. The field sizes were determined as the full width at half maximum. 
\newline
The flashDiamond reading was multiplied with a calibration factor, obtained by cross calibration with an Advanced Markus chamber in a clinical 18 MeV electron beam. An additional correction factor was used to correct for the difference in depth between the flashDiamond and the OSL sheet. This correction factor was determined as the ratio between the PDD at 15 mm depth with the one at 20 mm depth. The PDD curve was obtained by irradiation of a film, squeezed between RW3 plates, parallel with the beam. The film was read out and analyzed as described before, where a band of 5 mm was averaged.

\section{Results}
\subsubsection*{Calibration of the OSL system}
\begin{figure}[!ht]
    \centering
    \includegraphics[width=.8\textwidth]{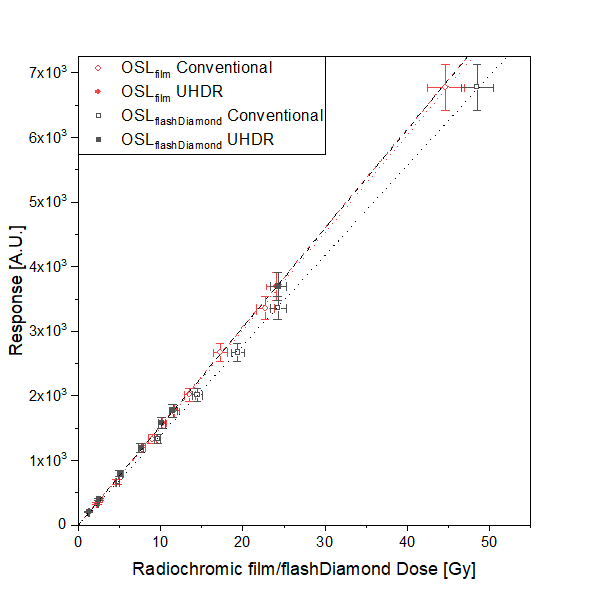}
    \caption{Calibration curves of the OSL sheet, and of the flashDiamond showing dose vs the radiochromic film dose in both conventional modality and UHDR modality.}
    \label{fig: Calibration}
\end{figure}

First, the dose rate was varied by switching between conventional and UHDR modalities. This implies a variation of the pulse amplitude. The OSL sheet was calibrated in both modalities and the effect of the pulse amplitude was investigated by comparison of the slopes. Figure~\ref{fig: Calibration} shows the calibration curves for the OSL sheet against grafchromic film (red) and the flashDiamond (grey). For both dosimeters, the calibration curves overlap within the standard deviation of the measurement. As the radiochromic film data consisted of a single measurement per data point, an uncertainty of 5\% was considered~\cite{marroquin2016evaluation}. The calibration coefficients were in agreement within 1.6\% and 9.5\% for calibration against radiochromic film and the flashDiamond respectively.

\subsubsection*{Pulse repetition frequency variation}
\begin{figure}[!ht]
    \centering
    \includegraphics[width=\textwidth]{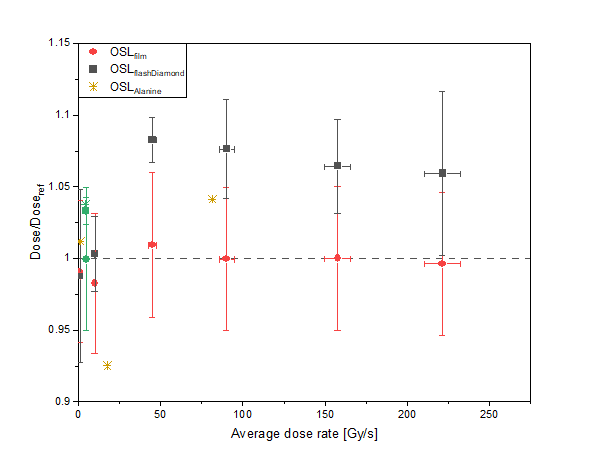}
    \caption{Normalized dose as function of dose rate for the OSL system and for flashDiamond. The filled and crossed symbols denote normalization with the radiochromic film and alanine dose measurements respectively. The green colored symbols represent the data points that were acquired with the same pulse length, PRF and SSD as the UHDR calibration. Dose rate was varied by changing the pulse repetition frequency.}
    \label{fig: PRF}
\end{figure}

The dose rate dependence of the OSL system was investigated by varying the pulse repetition frequency. The OSL dose readings, normalized with the flashdiamond and radiochromic film dose readings are shown for various PRF in Figure~\ref{fig: PRF}. After normalization to the flashDiamond, the OSL system shows an increasing dose response with average dose rate between 1.0 and 42.7 Gy/s. This results in a maximal difference of 10.5\%. After the initial increase, the dose response remains stable with average dose rate up to 212.1 Gy/s. However, when normalized to the radiochromic film, no dependence of the OSL dose response with average dose rate was observed, as reflected by the random distribution around 1.0 with a maximal deviation of 2.7\%. All data points fall within 1 sigma. The 3 additional measurements with alanine support the average dose rate independence as the normalized OSL dose response is spread around 1.0. It should be noted, however, that a maximal deviation of 7.5\% was observed for an average dose rate of 17.6 Gy.

\subsubsection*{Pulse length variation}
\begin{figure}[!ht]
    \centering
    \includegraphics[width=\textwidth]{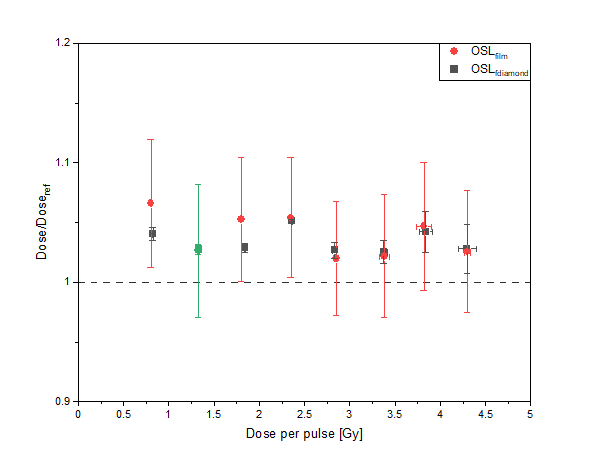}
    \caption{Normalized dose as function of dose per pulse for the OSL system and for the flashDiamond. The filled and crossed symbols denote normalization with the radiochromic film and alanine dose measurements respectively. The green colored symbols represent the data points that were acquired with the same pulse length, PRF and SSD as the UHDR calibration. Dose per pulse was varied by changing the pulse length.}
    \label{fig: PulseLenght}
\end{figure}

Next, the dose rate was varied by changing the pulse length, and thus the dose per pulse, in UHDR modality. Figure~\ref{fig: PulseLenght} shows the OSL dose readings, normalized by the flashDiamond and radiochromic film, for doses per pulse ranging from 0.8 up to 4.3 Gy. The data points, normalized to the flashdiamond are spread around 1.03, with a maximal difference of 2.5\% and an overlap within 3 sigma. The data points, normalized to radiochromic film are spread around 1.04, with a maximal difference of 4.4\% and all data points within 1 sigma from each other.  

\subsubsection*{Pulse amplitude variation}
Finally, the instantaneous dose rate was varied by variation of the SSD. Similarly, as in the first experiment, this also varies the pulse amplitude, but in a more gradual way. The response of the OSL system, normalized with the flashDiamond dose reading is randomly distributed around 1.02 for an instantaneous dose rate between 0.32 to 1.33 MGy/s. A maximum difference of 4.4\% is observed, with all data points falling within 1 sigma. The response of the OSL system, normalized with the radiochromic film dose reading, is randomly distributed around 1.05 with a maximum difference of 3.9\% and all data points falling within 1 sigma. 

\begin{figure}[!ht]
    \centering
    \includegraphics[width=\textwidth]{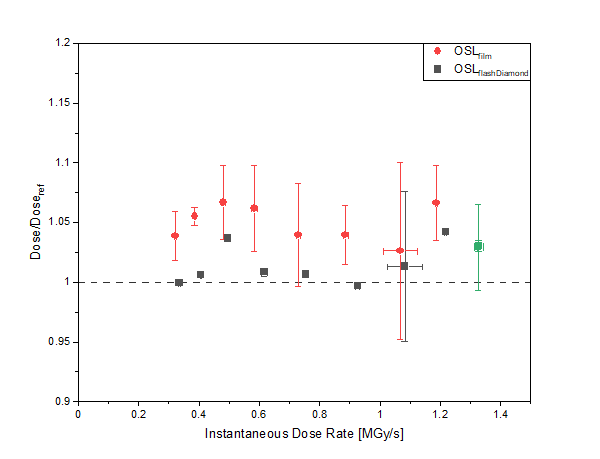}
    \caption{Normalized dose as function of instantaneous dose rate for the OSL system and for the flashDiamond. The filled and crossed symbols denote normalization with the radiochromic film and alanine dose measurements respectively. The green colored symbols represent the data points that were acquired with the same pulse length, PRF and SSD as the UHDR calibration. Instantaneous dose rate was varied by changing the source to surface distance.}
    \label{fig: iDR}
\end{figure}

\subsubsection*{2D characteristics}
\begin{figure}[!ht]
    \centering
    \includegraphics[width=\textwidth]{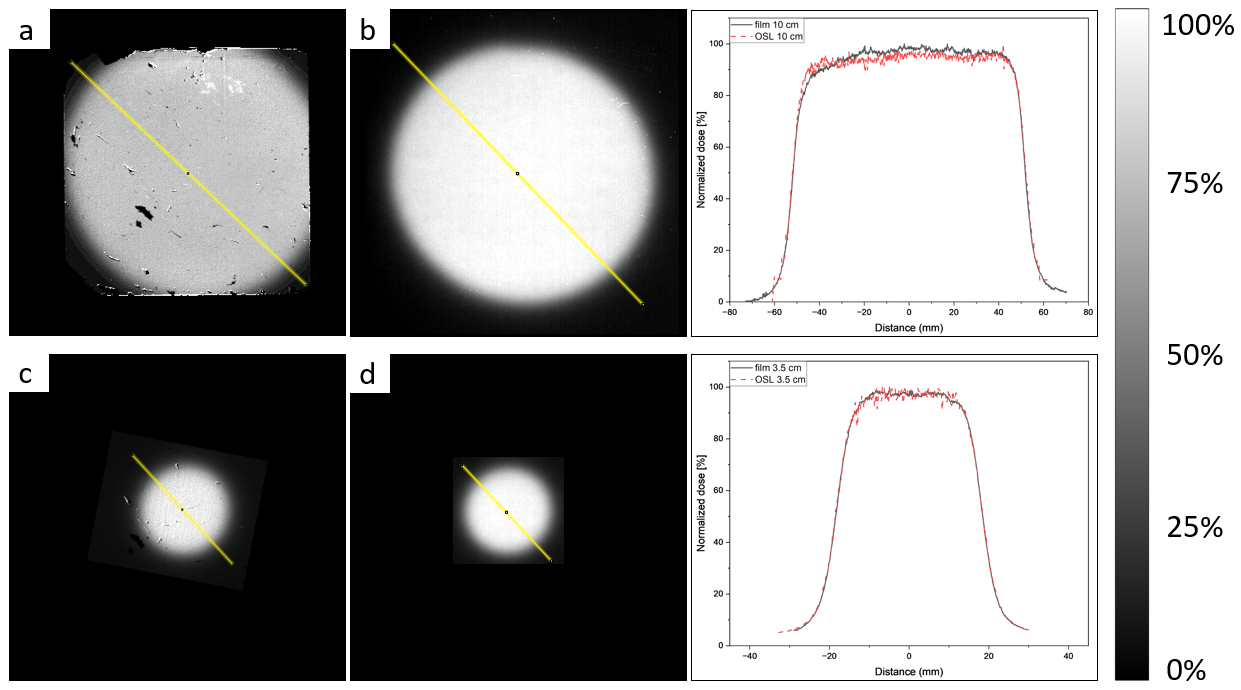}
    \caption{The normalized 2D dose maps with their respective profiles for the OSL system (red dashed line) and the radiochromic film (grey line) after irradiation with a 10 (a), (b) and 3.5 cm (c), (d) diameter field  respectively. The yellow denotes the pixels that were included to construct the profile.}
    \label{fig: Profiles}
\end{figure}

Representative 2D dose maps with their line profiles are shown in figure~\ref{fig: Profiles} for the radiochromic film and OSL system respectively. The yellow band in the images denotes the position of the line profile. The field size measured with radiochromic film are 10.35 $\pm$ 0.01 and 3.68 $\pm$ 0.01 cm for the nominal field sizes of 10 and 3.5 cm respectively. The field sizes measured with the OSL system for the same nominal field sizes are 10.41 $\pm$ 0.01 and 3.7 $\pm$ 0.01 cm respectively.

\newpage
 
\section{Discussion}
\begin{figure}[!ht]
    \centering
    \includegraphics[width=\textwidth]{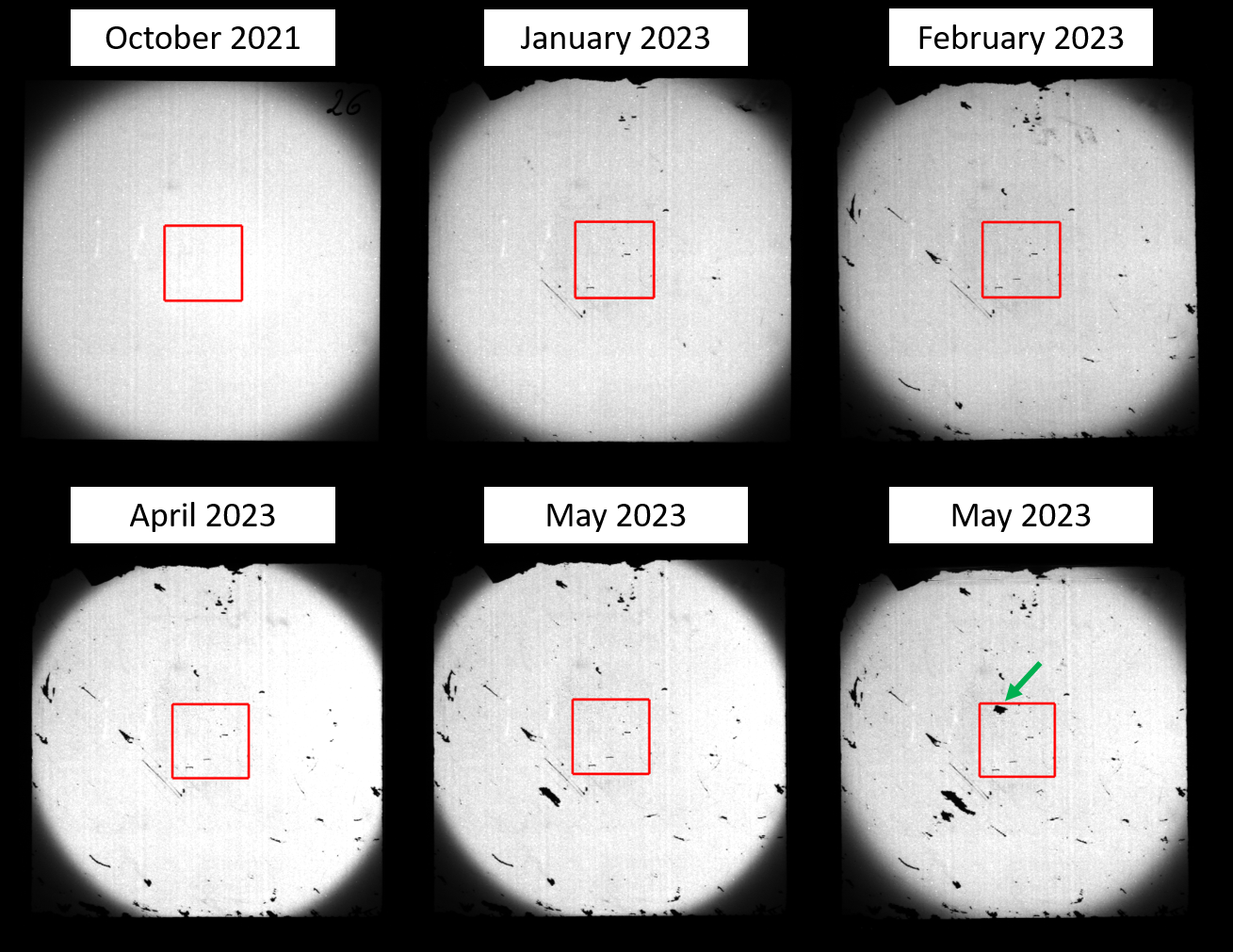}
    \caption{The OSL sheet irradiated using the same beam parameters at different time points. The extensive use and friction during positioning introduces scratches and patches where the coating got lost. The green arrow denotes a large patch within the region of interest that leading to a 6.0\% reduction of the dose reading.}
    \label{fig: OSLwear}
\end{figure}

The dose rate dependency of a 2D OSL system was investigated in an UHDR electron beam. One of the benefits of this system is the reusability of its sheet. However, the intensive use of the sheet and sliding it between RW3 plates many times, resulted in patches with no coating as can be seen in figure~\ref{fig: OSLwear}. As the dose rate dependency was investigated based on the average response within a ROI, the inclusion of such a patch within the ROI slightly affects the calibration. Therefore, the system was recalibrated for the average dose rate and dose per pulse experiment since the normalized dose readings showed a systematic offset of more than 10\%. 
It should be noted that a research sample was used. This implies that its coating is less homogeneous and robust than a final production coating. In addition, the risk of damage in a clinical environment is less, which increases the long term stability and life span of a single sheet. This statement is supported by the fact that previous studies with a similar coating, performed in a more clinical setting, did not lead to damage of the sheet. Hence hardness of the coating was not reported as a limiting factor~\cite{crijns2015t,wouter2017reusable,caprioli2022calibration}. A typical difference in setting is the direction of the beam. A clinical beam is most often vertically aligned, which allows more gentle handling when squeezing the OSL sheet between 2 RW3 slabs. Also, no tape is needed to keep the sheet in place. Therefore, the damage resulting from applying and removing the tape for every measurement can be avoided.

\subsubsection*{Calibration of the OSL system}
The difference of 1.6\% between the calibration factors for conventional and UHDR modality, obtained with radiochromic film, is small. Especially when considering that the dose uncertainty of radiochromic film is about 5\%. In addition, the conventional calibration curve falls within 1 sigma of the UHDR data points and vice versa. This observation makes it reasonable to assume that variation of the pulse amplitude, by switching the modality, does not affect the dose assessment of the OSL system. However, the calibration against the flashDiamond weakens this statement as a difference of 9.5\% is observed between conventional and UHDR calibration factors. We address this larger difference to the uncertainty associated with the depth correction of the flashDiamond.
\newline
This finding has an important consequence as it means that the calibration factor can be determined in the conventional beam, after which the OSL system can be used for dose assessment in the UHDR beam. Therefore, the OSL system can be cross calibrated against a clinical dosimeter, allowing traceability to a primary standard. A similar procedure is already generally accepted for radiochromic film in UHDR experiments. This result shows that the OSL system can be considered as at least equivalent to radiochromic film system when switching between modalities. 

\subsubsection*{Pulse repetition frequency variation}
The OSL dose, normalized to radiochromic film, is randomly spread around 1.0. This means that the radiochromic film and OSL system read the same dose, within the uncertainty. For the dose reading, normalized to the flashDiamond, this only holds true for an average dose rate up to 18.0 Gy/s. As there is a discrepancy between both reference detectors, 3 additional measurements with alanine dosimeters were performed. The OSL dose measurement, normalized to the alanine dose measurement shows a random distribution around 1, and, therefore, supports the average dose rate independence of the OSL system. The spread is however much larger compared to the radiochromic film normalized data with a maximal difference of 7.5\% from unity. As only a single measurement per point was performed, no uncertainties were included. However, comparing with the uncertainties of the data normalized to radiochromic film, it is reasonable to assume that the OSL dose readings, normalized by alanine, are within 1 sigma of 1.0. Therefore, we can conclude that the OSL system is at least equivalent to radiochromic film in terms of average dose rate dependence up to an average dose rate of 225 Gy/s with a PRF of 245 Hz. 
\newline
This reasoning also means that our flashDiamond detector is only stable up to an average dose rate of 18.0 Gy/s, above which the output is decreased. This behaviour is unexpected since previous studies have shown the PRF independence of the flashDiamond~\cite{marinelli2022design}. As more and more groups rely on flashDiamond and did not report such behavior, it is most probable that this is inherent to the specific prototype used here. This observation shows the relevance of this work and the difficulties in selecting a trustworthy, accurate reference dosimeter with low uncertainty in UHDR. Further investigation of this result is needed and ongoing with both the developers at the University of Rome Tor Vergata, and the manufacturer (PTW).

\subsubsection*{Pulse length variation}
The normalized dose measured with the OSL system suggests no dose per pulse dependency as all data points are within 1, and 3 sigma of each other for the radiochromic film and flashdiamond normalized doses, respectively. However, a systematic shift of, respectively, 3.9\%  and 3.4\% in output can be seen. As the shift is similar for both reference detectors, it is most likely due to the calibration of the OSL sheet. Indeed, comparing the obtained grey values of the calibration with the data point irradiated with the same beam parameters shows a decreased gray value of 6.0\%. This is because the calibration was acquired only after the pulse length variation experiment and an additional patch of coating damaged, as shown by the green arrow in figure~\ref{fig: OSLwear}. This reduction in average grey value leads to a higher calibration factor, and overestimation of the dose.

\subsubsection*{Pulse amplitude variation}
The normalized OSL dose readings in this experiment, suggest no dependency on instantaneous dose rate. All data points are randomly distributed around  1.02 and 1.05 for the dose readings normalized by the flashDiamond and radiochromic film, respectively. Also here a systematic shift is observed. The same reasoning as for the previous experiment is valid. The magnitude of the shift is with 4.4\% and 3.9\% for the dose readings normalized by the flashDiamond and radiochromic film, respectively, also similar as in the previous experiment.

\subsubsection*{2D characteristics}
After flat field correction, the profiles obtained with the OSL system are in good agreement with the ones obtained with radiochromic film. The OSL profile is slightly more noisy, which is due to a combination of the sheet uniformity and processing. The division of two images after registration, which requires interpolation, inherently increases the noise~\cite{wouter2017reusable}.  However, it should be considered that this OSL sheet was coated on a research line, inherently leading to a lower uniformity than with a production line. In addition, the flat field correction requires a registration of the flat field and the profile. This registration can be improved as it was performed manually. An automated registration could also further reduce the noise. The radiofchromic film overestimates the field size with 3.5\% and 5.1\%  for the 10 and 3.5 cm nominal field diameters, respectively. For the OSL system, similar overestimations of 4.1\% and 5.7\%, respectively are observed. This overestimation can be explained by the inclusion of a build up. The nominal field size is determined as the inner diameter of the applicator. However, the beam diverges when exiting the applicator. After passing the build up of 15 mm, the diameter will, therefore, be larger. For the 10 cm nominal field size, the radiochromic film and OSL are in agreement within 3 sigma, whereas for the 3.5 cm nominal field size the agreement is within 1 sigma. The overlap of the profiles and field size determinations suggest that the OSL system has the potential to have equivalent 2D characteristics as radiochromic film. However, to fully exploit this potential, both the coating and processing need optimization. The former includes improvement of the homogeneity and hardness of the coatings, which can be achieved by coating on a production line. The latter includes the introduction of 2D calibration and automation of the flat field correction. In addition, the definition of a well defined position in the carrier would reduce the noise and improve the reproducibility of the system.

\subsubsection*{Further perspectives}
One of the main limitations of the OSL system in the current study is the instability of the calibration. We showed that this is caused by damaging the coating during use. Since a production sample has improved scratch-resistance and uniformity, a similar study using a production sheet should be performed. An optimized phantom should be used as the vertical orientation of the sheet makes it more prone to coating damage. More important however, is the introduction of a pixel-wise dose-response calibration. This requires a - preferably automated - procedure to bring all the images in register. In addition, an optimization of the sheet position on the carrier could be considered to reduce the noise. 
\newline
An automated processing pipeline should be introduced and tested in a future study. Also, an uncertainty budget should be set up and compared with the one of radiochromic film, which is currently the used dosimeter for 2D reference dosimetry in UHDR beams. The 2D characteristics of the OSL system should be thoroughly investigated, and compared with radiochromic film with a gamma analysis. Finally, the energy independence and dark decay model for the OSL system should be investigated in an UHDR beam.

\section{Conclusions}
With the introduction of the possibly revolutionary FLASH-RT came the urgent the need for UHDR dosimetry. Accurate and fast dose distribution measurement and beam quality assurance is crucial for (pre)clinical studies. To date radiochromic film is considered the standard, but the drawback of cumbersome, lengthy readout protocols initiated the search for a valid alternative. In this work we performed a first characterization with respect to the main UHDR-related requirements for the 2D OSL system presented by Crijns et al.~\cite{wouter2017reusable}. We calibrated the OSL sheet for both modalities and investigated its dose rate dependency by variation of the average dose rate via the pulse repetition frequency, the dose per pulse via the pulse length, and the instantaneous dose rate via thedose rate modality and SSD. \newline
We showed that the calibration coefficient of the OSL sheet is interchangeable between conventional and UHDR modality. This allows to cross-calibrate the sheet against a clinical dosimeter in a conventional beam and allow traceability to a primary standard, even for UHDR irradiations. We showed the independence of the OSL system to average dose rate, pulse length and instantaneous dose rate. However, we saw systematic deviations in the pulse length and instantaneous dose rate experiments which are due to an overestimation of the calibration factor, due to physical damage of the coating. Despite the good results obtained in previous work, our prototype of the flashDiamond did show a dose rate dependence when varying the PRF. This shows the relevance and need for in-depth UHDR dosimetry studies.\newline
We showed that the 2D characteristics of the OSL system, in terms of field size determination, were equivalent with radiochromic film.\newline
These results demonstrate the promising nature of the OSL system to serve as an alternative for radiochromic film for reference dosimetry in UHDR electron beams. However, in order to confirm this statement, the hardness and uniformity of the coating should be improved, and automation and pixel-wise calibration should be introduced. Further in-depth characterization is needed to improve the robustness, accuracy and 2D properties of this system, in order to replace radiochromic film as 2D reference dosimeter in UHDR electron beams. 

\section{Acknowledgments}
This work was supported by VLAIO via the Flanders.HealthTech call [HBC.2021.0946]. SCK CEN and Iridium Network are also non-funded collaborators to the 18HLT04 UHDpulse project which received funding from the EMPIR programme. UHDR research in the Iridium Network is partially funded by Sordina IORT Technologies S.p.A..

\newpage
\printbibliography

\newpage
\appendix
\setcounter{table}{0}
\renewcommand{\thetable}{A\arabic{table}}
\section{Beam parameters}
\label{sec: appendix}
\begin{table}[!ht]
    \centering
    \caption{The beam parameters used for all experiments. The average dose rate, dose per pulse and instantaneous dose rate values were experimentally obtained with radiochromic film.}
    \resizebox{\textwidth}{!}{ 
    \begin{tabular}{|c|c|c|c|c|c|c|c|c|c|c|c| }
    \hline
         \multirow{2}{*}{Experiment} & \multirow{2}{*}{Modality} &  Energy & Field size & Pulse repetition & pulse length  & source surface & Number of  & Alanine & Average dose & Dose per & Instantaneous dose \\
         & &[MeV] & $\diameter$[cm] & frequency [Hz] & [$\mu$s] & distance [cm] &pulses& validation & rate [Gy/s] & pulse [Gy] & rate [MGy/s]\\
         \hline
         \multirow{14}{*}{Calibration} & Conventional & 9 & 3.5 & 5 & 1 & 68.2 & 12 & No & 0.95 & 0.19 & 0.19 \\
          \cline{2-12}
          & Conventional & 9 & 3.5 & 5 & 1 & 68.2 & 24 & No & 0.97 & 0.20 & 0.20 \\
          \cline{2-12}
          & Conventional & 9 & 3.5 & 5 & 1 & 68.2 & 48 & Yes & 0.94 & 0.19 & 0.19 \\
          \cline{2-12}
          & Conventional & 9 & 3.5 & 5 & 1 & 68.2 & 72 & No & 0.94 & 0.19 & 0.19 \\
          \cline{2-12}
          & Conventional & 9 & 3.5 & 5 & 1 & 68.2 & 96 & No & 0.90 & 0.18 & 0.18 \\
          \cline{2-12}
          & Conventional & 9 & 3.5 & 5 & 1 & 68.2 & 120 & No & 0.95 & 0.19 & 0.19 \\
          \cline{2-12}
          & Conventional & 9 & 3.5 & 5 & 1 & 68.2 & 240 & Yes & 0.93 & 0.19 & 0.19 \\
          \cline{2-12}
          & UHDR & 9 & 10 & 5 & 1 & 106 & 1 & No & 6.22 & 1.24 &  1.24  \\
          \cline{2-12}
          & UHDR & 9 & 10 & 5 & 1 & 106 & 2 & No & 6.89 & 1.38 & 1.38 \\
          \cline{2-12}
          & UHDR & 9 & 10 & 5 & 1 & 106 & 4 & No & 6.52 & 1.30 & 1.30 \\
          \cline{2-12}
          & UHDR & 9 & 10 & 5 & 1 & 106 & 6 & No & 6.50 & 1.30 & 1.30 \\
          \cline{2-12}
          & UHDR & 9 & 10 & 5 & 1 & 106 & 8 & No & 6.39 & 1.28 & 1.28 \\
          \cline{2-12}
          & UHDR & 9 & 10 & 5 & 1 & 106 & 9 & Yes & 6.48 & 1.30 & 1.30 \\
          \cline{2-12}
          & UHDR & 9 & 10 & 5 & 1 & 106 & 19 & No & 6.33 & 1.26 & 1.26\\
          \hline
          \multirow{7}{*}{PRF} & UHDR & 9 & 10 & 1 & 1 & 106 & 10 & Yes & 1.00 & 1.00 & 1.00  \\
         \cline{2-12}
          & UHDR & 9 & 10 & 5 & 1 & 106 & 10 & No & 4.84 & 0.97 & 0.97 \\
          \cline{2-12}
          & UHDR & 9 & 10 & 10 & 1 & 106 & 10 & Yes & 10.08 & 1.01 & 1.01 \\
          \cline{2-12}
          & UHDR & 9 & 10 & 50 & 1 & 106 & 10 & No & 45.02 & 0.90 & 0.90 \\
          \cline{2-12}
          & UHDR & 9 & 10 & 100 & 1 & 106 & 10 & Yes & 90.15 & 0.90 & 0.90 \\
          \cline{2-12}
          & UHDR & 9 & 10 & 175 & 1 & 106 & 10 & No & 157.68 & 0.90 & 0.90 \\
          \cline{2-12}
          & UHDR & 9 & 10 & 245 & 1 & 106 & 10 & No & 221.43 & 0.90 & 0.90 \\
          \hline
          \multirow{8}{*}{Pulse length} & UHDR & 9 & 10 & 5 & 0.5 & 106 & 16 & Yes & 4.02 & 0.80 & 1.61\\
         \cline{2-12}
          & UHDR & 9 & 10 & 5 & 1 & 106 & 9 & Yes & 7.46 & 1.49 & 1.49\\
          \cline{2-12}
          & UHDR & 9 & 10 & 5 & 1.5 & 106 & 6 & No & 9.39 & 1.88 & 1.25\\
          \cline{2-12}
          & UHDR & 9 & 10 & 5 & 2 & 106 & 4 & No & 11.75 & 2.35 & 1.17 \\
          \cline{2-12}
          & UHDR & 9 & 10 & 5 & 2.5 & 106 & 3 & No & 14.25 & 2.85 & 1.14 \\
          \cline{2-12}
          & UHDR & 9 & 10 & 5 & 3 & 106 & 3 & Yes & 16.91 & 3.38 & 1.13 \\
          \cline{2-12}
          & UHDR & 9 & 10 & 5 & 3.5 & 106 & 2 & No & 19.09 & 3.82 & 1.09 \\
          \cline{2-12}
          & UHDR & 9 & 10 & 5 & 4 & 106 & 2 & No & 21.50 & 4.30 & 1.08 \\
          \hline
           \multirow{8}{*}{Pulse amplitude} & UHDR & 9 & 10 & 5 & 1 & 111 & 10 & Yes & 5.94 & 1.19 & 1.19 \\
         \cline{2-12}
          & UHDR & 9 & 10 & 5 & 1 & 116 & 11 & No & 5.34 & 1.07 & 1.07\\
         \cline{2-12}
          & UHDR & 9 & 10 & 5 & 1 & 126 & 13 & No & 4.43 & 0.89 & 0.89\\
          \cline{2-12}
          & UHDR & 9 & 10 & 5 & 1 & 136 & 15 & Yes & 3.65 & 0.73 & 0.73 \\
          \cline{2-12}
          & UHDR & 9 & 10 & 5 & 1 & 146 & 17 & No & 2.92 & 0.58 & 0.58 \\
          \cline{2-12}
          & UHDR & 9 & 10 & 5 & 1 & 156 & 19 & Yes & 2.40 & 0.48 & 0.48 \\
          \cline{2-12}
          & UHDR & 9 & 10 & 5 & 1 & 166 & 22 & No & 1.87 & 0.37 & 0.37 \\
          \cline{2-12}
          & UHDR & 9 & 10 & 5 & 1 & 176 & 25 & No & 1.61 & 0.32 & 0.32 \\
          \hline
    \end{tabular}}
    \label{tab: Beam parameters}
\end{table}

\end{document}